\documentclass[%
 prb,
 jmp,%
 amsmath,amssymb,
 reprint,%
  superscriptaddress,
]{revtex4-1}
\usepackage[english]{babel}
\usepackage[latin1]{inputenc}
\usepackage{graphicx}
\usepackage{subfigure}
\usepackage{epstopdf}
\usepackage{dcolumn}
\usepackage{bm}
\usepackage{stackrel}
\draft 

\begin{document}


\title{Theory of second harmonic generation in few-layered MoS$_2$} 


\author{Mads L. Trolle}
\email[]{mlt@nano.aau.dk}
\affiliation{Department of Physics and Nanotechnology, Aalborg University, Skjernvej 4A, DK-9220 Aalborg East, Denmark}
\author{Gotthard Seifert}
\affiliation{Physikalische Chemie, Technische Universität Dresden, D-01062 Dresden, Germany}
\author{Thomas G. Pedersen}
\affiliation{Department of Physics and Nanotechnology, Aalborg University, Skjernvej 4A, DK-9220 Aalborg East, Denmark}
\affiliation{Center for Nanostructured Graphene (CNG), Aalborg University, DK-9220 Aalborg East, Denmark}

\date{\today}
\begin{abstract}
Recent experimental results have demonstrated the ability of monolayer MoS$_2$ to efficiently generate second harmonic fields with susceptibilities between 0.1 and 100 nm/V. However, no theoretical calculations exist with which to interpret these findings. In particular, it is of interest to theoretically estimate the modulus of the second harmonic response, since experimental reports on this differ by almost three orders of magnitude. Here, we present single-particle calculations of the second harmonic response based on a tight-binding band structure. We compare directly with recent experimental findings and include in the discussion also spectral features and the effects of multiple layers.
\end{abstract}

\maketitle 
The prediction and observation of a direct band gap in monolayer (ML) MoS$_2$ has revitalized the interest in the optical properties of this material \cite{Molina2013,Mak2010,Splendiani2011,Kumar2013,Malard2013,Heinz2013,Eda2011,Yin2012,Sundaram2013}. Indeed, a substantial photoluminescence for 1H-MoS$_2$ has been observed\cite{Mak2010,Splendiani2011}. Additionally, several papers\cite{Kumar2013,Malard2013,Heinz2013} have recently demonstrated how second harmonic generation (SHG) microscopy can be used to extract important information regarding e.g. the number of layers and crystallographic orientation of few-layered MoS$_2$ platelets. Furthermore, exfoliated MoS$_2$ was shown experimentally to display a remarkably large second harmonic signal, with second harmonic susceptibilities on the order of $\sim 100 \ $ nm/V reported in Ref. \onlinecite{Kumar2013} while Refs. \onlinecite{Malard2013,Heinz2013} report only $\sim 0.1 \ $ nm/V  (assuming a homogeneous susceptibility inside the monolayer of thickness $\sim 3 \ \textrm{\AA}$ when relating to sheet second harmonic susceptibilities). Regions covered with an odd number of 2H stacked layers were found to generate second harmonic fields with efficiencies decreasing slightly with the number of layers, while regions with an even number of layers displayed almost vanishing second harmonic signals as expected due to centro-symmetry.\cite{Kumar2013,Malard2013,Heinz2013} However, few-layered MoS$_2$ grown by chemical vapour deposition does not follow this trend\cite{Kumar2013}, possibly due the stacking order of CVD grown films deviating from 2H. 
In Ref. \onlinecite{Malard2013}, second harmonic spectra of both ML and trilayer (TL) MoS$_2$ were presented, demonstrating an intense peak in the second harmonic spectrum at pump photon energies near 1.45 eV, with a slight redshift for TLs compared to MLs. 
\begin{figure}[hbtp]
\includegraphics[width=0.5\textwidth]{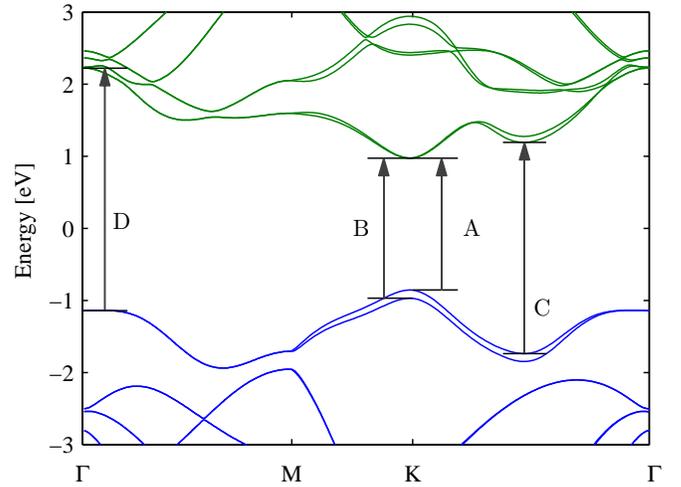}
\caption{Band structure of ML MoS$_2$.  Arrows indicate important optical transitions.}
\label{fig:BS}
\end{figure}

There exists no theoretical work, with which to interpret the experimental findings mentioned above. Hence, in this letter we consider the microscopic origins of the second harmonic response of few-layered MoS$_2$ based on a tight-binding $sp^3d^5$ band structure recently published\cite{Zahid2013}. 
%
%
At present, it is very computationally demanding to retain the k-space resolution needed to resolve delicate spectral features in a full exciton Bethe-Salpeter calculation. Moreover, the single particle results remain an important first step in the theoretical understanding of the nonlinear optical properties of MoS$_2$. We have therefore chosen to omit excitonic effects in the present work and employ the single-particle second harmonic response formalism developed by Moss and Sipe\cite{Ghahramani1991,TGP2009}. 
%
%
%
%
We verify that the experimental measurements in Refs. \onlinecite{Malard2013, Heinz2013} agree to within an order of magnitude with our model, and proceed to analyse the resonance structure of the second harmonic spectrum. We also analyse the dependence of $\chi^{(2)}$ on the number of 2H stacked layers and find little difference in the general magnitude of $\chi^{(2)}$ for varying odd-numbered layers, although slight changes to the spectral features are observed.
\begin{figure*}[hbtp]
\subfigure[]{
\includegraphics{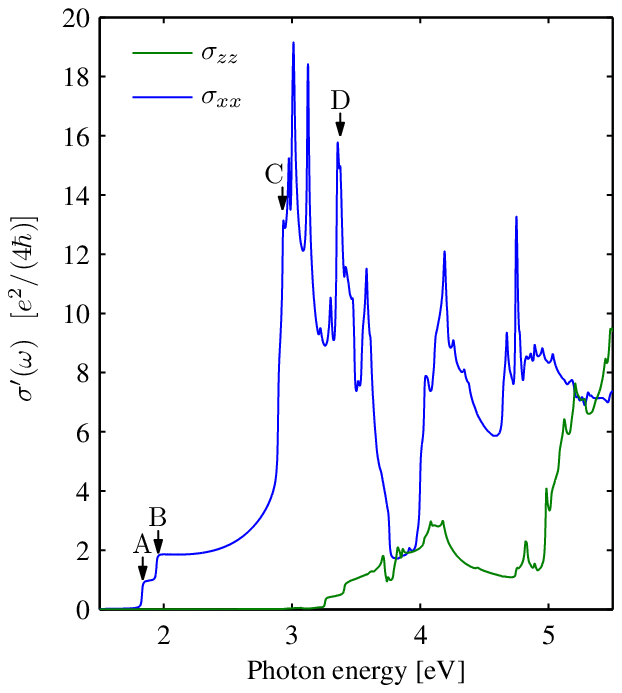}
\label{fig:linear}
}
\subfigure[]{
\includegraphics{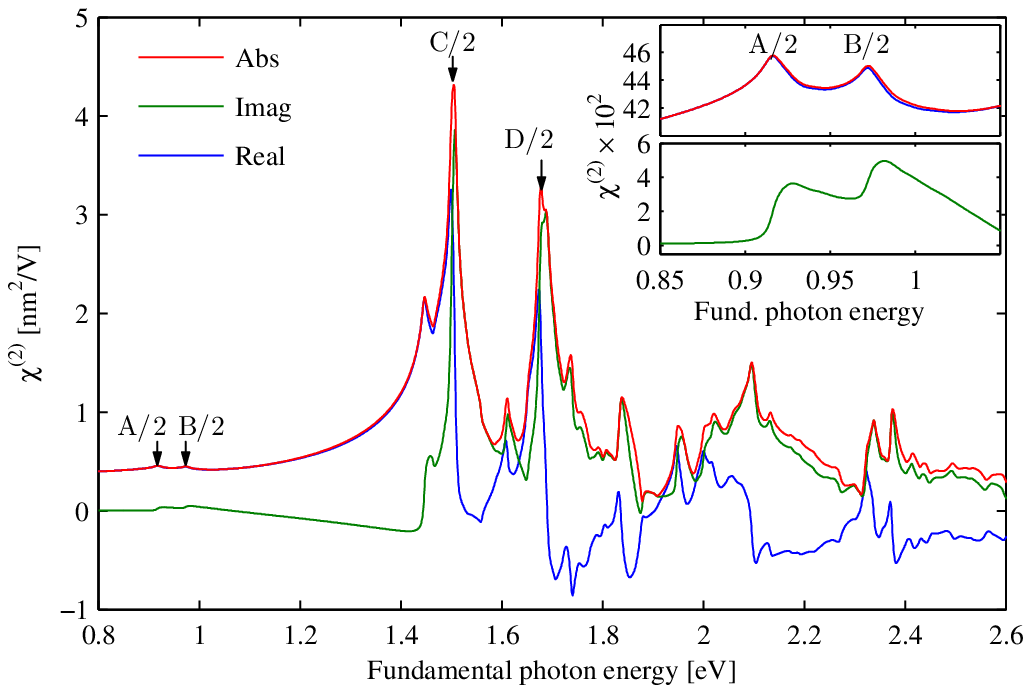}
\label{fig:SHG}
}
\caption{\subref{fig:linear} Linear optical sheet conductivity tensor of a MoS$_2$ ML, with the in-plane tensor elements denoted by $\sigma_{xx}=\sigma_{yy}$ whereas the out-of-plane response is $\sigma_{zz}$ . \subref{fig:SHG} Second harmonic sheet response of MoS$_2$ ML. The static response is $\chi^{(2)}(\omega=0) = 0.3$ nm$^2$/V.}
\end{figure*}

It is well known that the valence and conduction band extrema of ML MoS$_2$ are located at the K points of the Brillouin zone and are dominated by $d$-orbitals localized on the Mo atoms.\cite{Cappelluti2013} 
These bands are particularly important for optical transitions in the visible range, and are well represented by the tight-binding band structure of Ref. \onlinecite{Zahid2013}. However, a significant spin-orbit coupling, due to the heavy Mo atoms, causes a $\sim$ 100 meV splitting of the two highest valence bands near the K-points. To properly account for this we include spin-orbit coupling between $d$-orbitals localized on the same Mo atom\cite{Jones2009}, instead of the $p$-orbitals used in the original parametrization\cite{Zahid2013}. We fit the spin-orbit parameter $\lambda_{d,\textrm{Mo}}=54 \ \textrm{meV}$ to the 112 meV splitting of the two highest valence bands at K reported in Ref. \onlinecite{Molina2013}. 
The band gap of 1.8 eV generated using this method is comparable to the DFT band gap\cite{Zahid2013,Molina2013} of $1.7-1.8$ eV and the experimentally recorded optical absorption edge\cite{Malard2013,Frindt1965,Mak2010} of 1.9 eV. For this reason, we choose to neglect quasi-particle effects aiming instead to reproduce the position of experimental spectral features.
However, it should be noted that the agreement between the DFT band gap and the experimental absorption peak position arises from the approximate cancellation of the band gap increase due to quasi-particle (GW) corrections and the red-shift of optical features due to the exciton binding energy\cite{Molina2013,Komsa2012}.
%
%
%
The band structure of ML MoS$_2$ is displayed in Fig. \ref{fig:BS} together with some optical transitions to be discussed shortly.
We calculate the real part of the diagonal, linear optical sheet conductivity tensor using the well-known expression\cite{TGP2008_2}
\begin{align}
\sigma_{aa}^{\prime}(\omega) = \frac{e^2}{2\pi m^2 \hbar \omega} \sum_{c,v} \int |p_{cv}^{a}|^2 \delta(\omega-\omega_{cv}) \ d^2k, 
\end{align}
where  $\omega_{ij}$ and $p_{ij}^a$ denote, respectively, the transition frequency and the $a$-component of the momentum matrix element between states in bands $i$ and $j$ (with an implicit $k$-dependency). Momentum matrix elements are calculated as in Ref. \onlinecite{TGP2003}. Furthermore, the band indices $c$ and $v$ indicate conduction and valence bands, respectively. \newline
The imaginary part of the interband sheet second harmonic susceptiblity tensor at fundamental pump frequency $\omega$ can be calculated using\cite{Ghahramani1991,TGP2009}
\begin{align}
\chi_{abc}^{(2)\prime \prime}&(\omega) = \frac{e^3}{ 2 \pi m^3\hbar^2 \epsilon_0 \omega^3} \sum_{c,v,l}\int \left[ \frac{P_{vcl}}{\omega-\omega_{lv}}\delta(2\omega-\omega_{cv}) \right. \notag \\
 &\left.+ \left( \frac{P_{vlc}}{\omega+\omega_{cl}} + \frac{P_{clv}}{\omega+\omega_{lv}}  \right)\delta(\omega-\omega_{cv}) \right] d^2k. \label{eq:SHG}
\end{align}
Here, $P_{ijl} = \textrm{Im} \left\{ p_{ij}^a \left(p_{jl}^b p_{li}^c + p_{jl}^c p_{li}^b \right) \right\}/2$ and the band index $l$ runs over all bands with the restriction $l \neq (c,v)$.
The first term in Eq. \ref{eq:SHG} contributes when an electronic excitation frequency $\omega_{cv}$ resonant with the second harmonic photon frequency $2 \omega$ can be found. This term is referred to as the $2\omega$-term, and whenever  the aforementioned criterion is satisfied while $\omega \approx \omega_{lv}$ a particularly powerful, so-called double resonance is found. Similar comments can be made regarding the $\omega$-terms that contribute whenever the condition $\omega = \omega_{cv}$ is fulfilled.
Due to symmetry, the only non-vanishing second harmonic tensor elements are $\chi^{(2)}_{xxx} = -\chi^{(2)}_{xyy} = -\chi^{(2)}_{yyx} = - \chi^{(2)}_{yxy} \equiv \chi^{(2)}$, with the $x$-axis aligned along an armchair direction. Malard \textit{et al.}\cite{Malard2013} and Kumar \textit{et al.}\cite{Kumar2013} define this direction differently relative to the indices of the contributing tensor elements. However, we follow the conventions of Ref. \onlinecite{Malard2013} and confirm these to be correct by numerical testing (which is also clear from symmetry, since the armchair direction spans a mirror plane together with the $z$-axis). 
 The $k$-integrations are performed using the improved linear-analytic triangle method\cite{TGP2008_2}, where care is taken to analyse double resonances of Eq. \ref{eq:SHG} by subsequent refinement of the integration mesh. Having calculated the imaginary part of $\chi^{(2)}$, the real part is found by Kramers-Kronig transformation. We generally apply a phenomenological broadening of 5meV.

The calculated linear optical response for a ML is plotted in Fig. \ref{fig:linear}. The absorption edge is dominated by two step-like features in agreement with other single-particle results\cite{Molina2013} following from the relatively large splitting of the highest valence bands in ML MoS$_2$ due to spin-orbit interaction. 
Hence, transitions from the highest and second highest valence bands to the lowest conduction bands, as indicated in Fig. \ref{fig:BS}, give rise to the distinct A and B steps in the single-particle spectrum of Fig. \ref{fig:linear}.  It is noted that a full Bethe-Salpeter treatment enhances the A and B features into peaks red-shifted by 1.1 eV compared to the quasi-particle spectrum\cite{Komsa2012}, and in experiments these features are indeed observed as peaks\cite{Malard2013,Mak2010,Frindt1965}. At larger energies corresponding to the band gap on the $\Gamma$K-line, several transitions from nearly parallel bands contribute to the absorption. The first of these are indicated as the C transition in Fig. \ref{fig:BS}, while a structure of subsequent peaks follow from spin-obit splitting similarly to the A and B transitions. Collectively, these give rise to a broad peak upon inclusion of additional phenomenological broadening denoted the C transition. At photon energies corresponding to the band gap at $\Gamma$, an absorption peak arising from transitions indicated by D in Fig. \ref{fig:BS} is seen, with a similar peak structure.

\begin{figure}[hbtp]
\includegraphics[width=0.5\textwidth]{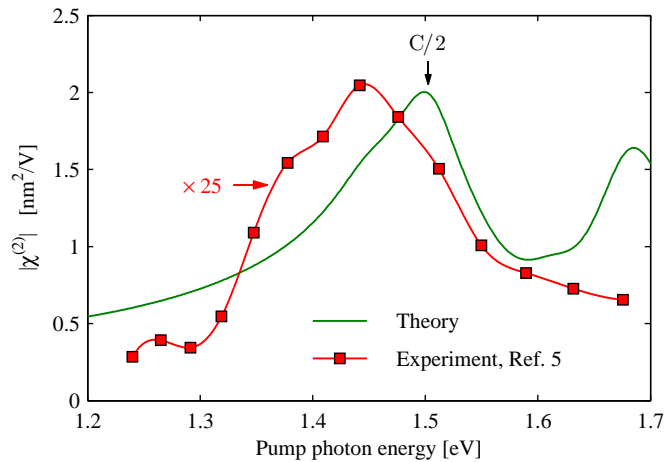}
\caption{Comparison between our theory and the experimental results of Ref. \onlinecite{Malard2013}. A phenomenological broadening of 25 meV has been applied to the theoretical results and the red line is a guide for the eye.}
\label{fig:exp}
\end{figure}
 The calculated second harmonic susceptibility of a MoS$_2$ ML is shown in Fig. \ref{fig:SHG}. The low-energy response near half the band gap is due to the $2\omega$-term of Eq. \ref{eq:SHG} only, and can be interpreted as optical transitions, for which the second harmonic photon energy matches transitions from the highest valence band to the lowest conduction band near the K-points. Two peaks are observed at fundamental photon energies corresponding to half the photon energies of the A and B peaks in the linear spectrum, although these are much weaker relative to the remaining spectrum when compared to the corresponding step heights in Fig. \ref{fig:linear}. Hence, we have included a magnified view of this spectral region as insets in Fig. \ref{fig:SHG}. These peaks correspond to $2\omega$-processes, and are here termed the A$/2$ and B$/2$ features since they appear at half the pump photon frequency of the comparable linear peaks. Similarly, one might expect peaks corresponding to $\omega$-processes at twice the fundamental photon energy. However, in this region $2\omega$-processes close to double resonance dominate making the $\omega$-features difficult to observe. 
 Also, an intense peak is found in the second harmonic spectrum at fundamental photon energies $\sim 1.45 \ $ eV corresponding to the half the energy of the C transition. We denote this feature C$/2$, and note that its spectral position fits very well with the experimentally observed SHG peak at 1.45 eV of Ref. \onlinecite{Malard2013}. 
 This can readily be seen in Fig. \ref{fig:exp}, where we compare our theoretical calculations (now including a broadening of 25 meV realistic for comparison with room-temperature experiments) with the spectrum recorded by Malard \textit{et al.}\cite{Malard2013}. A good agreement in peak position and shape is generally observed, however, a rather large intensity difference is also found. We stress that Kumar \textit{et al.}\cite{Kumar2013} report susceptibilities two order of magnitude larger than the those calculated by us, making it clear  that some uncertainties in the experimentally determined susceptibilities must be expected for an atomically thin material.  
 Also, we here compare theoretical results calculated for a free-standing MoS$_2$ sheet with experiments performed on a substrate where e.g., strain or substrate phonons may be important. 
 We also stress that excitonic effects are neglected in the current model, however, we believe that inclusion of these will have three main effects. Firstly, all spectral features arising from 2$\omega$ processes will be red-shifted by half the exciton binding energy, while features arising from $\omega$-processes will be red-shifted by the binding energy, compared to the quasi-particle spectrum (and, hence, remain at nearly the same spectral position compared to results derived directly from density-functional theory, as mentioned previously). Secondly, similarly to what is observed in the linear case, low energy features (such as the A$/2$ and B$/2$ peaks) will be enhanced into more powerful peaks at the expense of intensity at larger energies (such as the C$/2$ peak). Thus, this effect might explain part of the intensity difference between experiment and theory in Fig. \ref{fig:exp}. Thirdly, red-shifting excited states could introduce new double resonance conditions and, hence, change the second harmonic spectrum. The effects of the latter are difficult to predict without a full calculation including electron-hole interaction.
 Thus, while we expect some disagreement between theory and experiment in the observed peak intensity, we believe its position and shape are reproduced rather well. Additionally, we expect the order of magnitude of the off-resonance susceptibilities reported here to be realistic.
 %
 %
  %
 Interestingly, transitions corresponding to the weak shoulder on the C peak near 3 eV of the calculated linear spectrum results in a small peak near 1.5 eV if Fig. \ref{fig:SHG}. Upon including broadening,  this results in a shoulder near 1.45 eV in Fig. \ref{fig:exp}, which also appears in the experimental spectrum of the same figur.
 A peak in the second harmonic spectrum corresponding to the D transition is also observed in Fig. \ref{fig:SHG}. Furthermore, the static susceptibility was found to be $\chi^{(2)}(\omega=0) = 0.3 \ $ nm$^2$/V.

\begin{figure}[hbtp]
\includegraphics{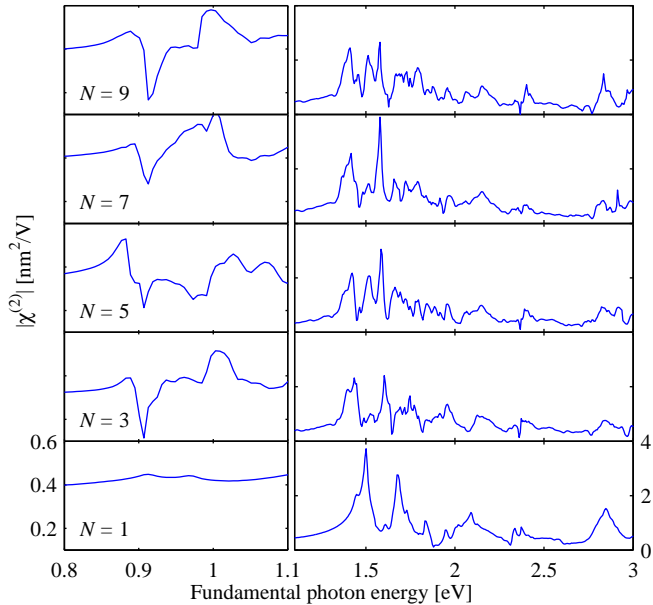}
\caption{Sheet second harmonic response of few-layered MoS$_2$ with $N$ layers. The scales on all boxes are identical.}
\label{fig:SHGFL}
\end{figure}
In Fig. \ref{fig:SHGFL}, we present results for second harmonic generation in few-layered MoS$_2$ with varying number $N$ of 2H-stacked layers. We observe that the general magnitude of the sheet second harmonic response remains unchanged with increasing $N$, while the C$/2$ feature is smeared out leading to an overall reduction in the on-resonance intensity of approximately 30 \% in good agreement with the experimental findings of Refs. \onlinecite{Kumar2013} and \onlinecite{Malard2013}. This trend can be understood by considering the weakly coupled MoS$_2$ layers; if these were completely decoupled, $N$ layers would contribute the same second harmonic polarization, but with phases alternating by $\pi$ due to adjacent layers being mirror images of each other in the $yz$-plane. 
Hence, for an odd number of layers, the net second harmonic polarization would equal the response of a single layer. The most important effect of introducing a weak interlayer coupling is to increase the splitting of the top valence bands at K, introducing slight changes to the second harmonic response near half the band gap at K. This causes modifications of the A$/2$ and B$/2$ features, while similar splitting throughout the Brillouin zone smears out the C$/2$ peak. While luminescence properties are obviously greatly affected by the transition of MoS$_2$ from a direct to an indirect gap semiconductor upon including multiple layers\cite{Mak2010,Splendiani2011}, this has little consequence for the properties studied here. 

In conclusion, we find clear signatures of the A and B excitations known from linear optics also in the second harmonic spectrum of ML MoS$_2$ at photon energies near half the band gap. With increasing number of layers, the spectral separation between the A and B peaks predictably increases with increased splitting of the two highest valence bands. Furthermore, a very intense peak corresponding to the C excitation in linear optics was found at second harmonic photon energies near the band gap on the $\Gamma$K-line in $k$-space. This peak position agrees very well with the experimentally observed peak\cite{Malard2013} at pump photon energies of 1.45 eV. The magnitude of the sheet second harmonic response was found to be $\sim$ 0.3 nm$^2$/V off resonance and up to $\sim$ 4 nm$^2$/V on resonance, placing it in between the values reported by Malard \textit{et al.}\cite{Malard2013} and Li \textit{et al}\cite{Heinz2013}  (agreeing to within an order of magnitude) and those reported by Kumar \textit{et al.}\cite{Kumar2013} (agreeing to within two orders of magnitude).
\section*{Acknowledgements}
TGP gratefully acknowledges the financial support from the Center for Nanostructured Graphene (project DNRF58) financed by the Danish National Research Foundation. Also, the authors gratefully acknowledge Ana Maria de Paula of the Federal University of Minas Gerais, Brazil, for supplying the experimental data of Fig. \ref{fig:exp} first published in Ref. \onlinecite{Malard2013}.

\bibliography{bib}

\end{document}